\documentclass[%
 reprint,
showpacs,
 amsmath,amssymb,
 aps,
 pra,
longbibliography,
floatfix
]{revtex4-1}

\usepackage[utf8]{inputenc}	
\usepackage[T1]{fontenc}	
\usepackage{fixltx2e,graphicx}		
\usepackage{color}
\usepackage[english]{babel}
\usepackage{hyperref}
\usepackage{url}
\usepackage{todonotes}
\usepackage{lipsum}
\usepackage{bm}
\usepackage{bbold}
\usepackage{mathtools}
\usepackage{braket}
\usepackage{textcomp}

\usepackage[position=top,caption=false]{subfig}

\let\originaleqref\eqref
\renewcommand{\eqref}{Eq.~\originaleqref}

\newcommand{\fref}[1]{\figurename~\ref{#1}}

\newcommand{\ac}{\hat{a}_c}
\newcommand{\acd}{\hat{a}_c^\dagger}
\newcommand{\dac}{\dot{\hat{a}}_c}
\renewcommand{\a}{\hat{a}}
\newcommand{\ad}{\hat{a}^\dagger}
\newcommand{\q}{\hat{q}}
\newcommand{\Sm}{\hat{S}_-}
\newcommand{\Sp}{\hat{S}_+}
\newcommand{\Sz}{\hat{S}_z}
\newcommand{\Sx}{\hat{S}_x}
\newcommand{\Sy}{\hat{S}_y}
\newcommand{\Sznh}{S_z}
\newcommand{\Sxnh}{S_x}
\newcommand{\Synh}{S_y}
\newcommand{\dSm}{\dot{\hat{S}}_-}

\newcommand{\dSz}{\dot{\hat{S}}_z}
\newcommand{\dSx}{\dot{\hat{S}}_x}
\newcommand{\dSy}{\dot{\hat{S}}_y}

\renewcommand{\d}[0]{\hat{d}}
\newcommand{\dd}[0]{\hat{d}^\dagger}
\renewcommand{\b}[0]{\hat{b}}
\newcommand{\bd}[0]{\hat{b}^\dagger}

\newcommand{\X }[0]{{\hat{X}}}
\newcommand{\Y }[0]{{\hat{Y}}}
\newcommand{\Xnh }[0]{{X}}
\newcommand{\Ynh }[0]{{Y}}
\newcommand{\dX }[0]{\dot{\hat{X}}}
\newcommand{\dY }[0]{\dot{\hat{Y}}}
\renewcommand{\H }[0]{\hat{H}}
\newcommand{\C}[3]{C_{#1, #2{\rightleftharpoons} #3} }

\newcommand{\e}[1]{\mathrm{e}^{#1}}

\newcommand{\tp}[0]{{t^\prime}}

\newcommand{\sx}[0]{\hat{\sigma}_\mathrm{x}}
\newcommand{\sy}[0]{\hat{\sigma}_\mathrm{y}}
\newcommand{\sz}[0]{\hat{\sigma}_\mathrm{z}}
\renewcommand{\sp}[0]{\hat{\sigma}_\mathrm{+}}
\newcommand{\sm}[0]{\hat{\sigma}_\mathrm{-}}

\graphicspath{
{../../Plots/}
{../../plots/}
}

\begin{document}
\title{Relaxation of an ensemble of two-level emitters in a squeezed bath}
\author{Alexander Holm Kiilerich}
\email{kiilerich@phys.au.dk}
\author{Klaus Mølmer}
\email{moelmer@phys.au.dk}

\date{\today}
\affiliation{Department of Physics and Astronomy, Aarhus University, Ny Munkegade 120, 8000 Aarhus C, Denmark}
\date{\today}

\bigskip

\begin{abstract}
We derive and evaluate equations of motion for the mean values and variances of the components
of spins collectively  coupled to a broadband squeezed radiation reservoir. Our formalism bridges between a single
two-level emitter, represented by spin components that relax at different rates,  depending on the degree of squeezing,
to an ensemble of emitters represented by a large collective spin, whose components relax independently of the squeezing.
For a single spin, the steady-state fluctuations in the transverse components are independent of the squeezing, while the steady state of a large spin ensemble reflects the statistics of the squeezed reservoir. This follows from an analysis of the Langevin noise contributions to the equations of motion and their consequences for the first and second moments of the spin operators. We argue that the difference between a single and many spins is related to whether vacuum fluctuations or radiation reaction dominate the coupling of the spin system to the radiation environment.
\end{abstract}

\maketitle
\noindent

\section{Introduction} \label{sec:intro}
The interaction between an atom and the quantized radiation field played a defining role in the early formulation of quantum mechanics. It constitutes the basis for absorption and emission of light, and it causes relaxation and decoherence in otherwise isolated material quantum systems. Quantum mechanics imposes uncertainty relations on non-commuting observables and when quantum theory is applied to electromagnetic fields, even the zero-photon vacuum state is equipped with electromagnetic amplitude fluctuations. While such fluctuations appear to pose a fundamental limit to, e.g., optical measurements, the development of sources producing squeezed states of light (see, e.g., \cite{andersen201630,PhysRevLett.104.251102,yurke1989squeezing,PhysRevB.89.214517}) allows higher sensitivity probing of the squeezed field quadrature \cite{Giovannetti19112004}.

Material quantum systems have quantum degrees of freedom such as position, momentum, angular moment, and spin components with well-known uncertainty relations. Since such systems are routinely being employed in precision tests and measurements, it is an attractive possibility to squeeze the relevant degrees of freedom or to squeeze the light used to probe the system in an experiment. Examples include optomechanical devices and ensembles of atoms, molecules, and spin dopants in solids, studied for, e.g., gravitational wave detection \cite{ligo2011gravitational}, atomic clocks \cite{PhysRevA.50.67}, and fundamental tests \cite{PhysRevLett.93.020401}. Furthermore, squeezed light has been demonstrated to provide improved sensitivity in spectroscopy setups \cite{PhysRevLett.68.3020,PhysRevA.93.053802}, in biological particle tracking \cite{taylor2013biological}, and recently in magnetic resonance detection \cite{PhysRevX.7.041011}. Generally, squeezing in multi-particle and multimode optical systems entails entanglement, allowing squeezed states of both light and matter to find additional applications in quantum information protocols such as  teleportation \cite{PhysRevLett.80.869,furusawa1998unconditional,RevModPhys.77.513}.

If squeezed radiation is used to probe a quantum system, one must take the influence exercised by the field on that system into account. For example, when the fluctuations in the electromagnetic field modes are altered, so are the derived relaxation dynamics of the systems interacting with the field. Gardiner's seminal result on a single two-level system in a squeezed reservoir \cite{PhysRevLett.56.1917} shows how the decay
of the excitation and the two transverse Bloch vector components is modified by the coupling to that reservoir. This effect has recently been observed in a super conducting qubit coupled to a squeezed vacuum microwave field produced by a Josephson parametric amplifier \cite{murch2013reduction,PhysRevX.6.031004}, and it is equivalently witnessed in the resonance fluorescence spectrum \cite{PhysRevLett.58.2539} and the steady-state inversion of a driven system \cite{PhysRevA.46.4354}.
A single two-level system can be represented as a spin-1/2 particle with a spin-up (excited) and a spin-down (ground) state.
In the Holstein-Primakoff approximation (HPA) \cite{PhysRev.58.1098}, an ensemble of weakly excited spins is equivalent to an oscillator mode. An oscillator experiences a squeezed reservoir as Langevin noise acting asymmetrically on its two quadratures but unlike a spin-1/2 system, the relaxation of the quadratures is not affected by the squeezing, which instead manifests itself directly in their variances.

The interaction of an ensemble of atoms with a broadband
squeezed reservoir of field modes has been studied theoretically
with the master equation formalism (see, e.g., \cite{dalton1999atoms}) and it has been shown that the collective interaction of two-level atoms with squeezed vacuum fields, which have minimum-uncertainty product of the field quadratures, leads to a pure steady state for a pair \cite{PhysRevA.39.1962} and for any even number of spins \cite{PhysRevA.41.3782}. These steady states are spin squeezed with minimum uncertainty product of their squeezed and anti-squeezed  spin components \cite{PhysRevA.41.3782}.

\begin{figure}
\includegraphics[trim=0 0 0 0,width=0.9\columnwidth]{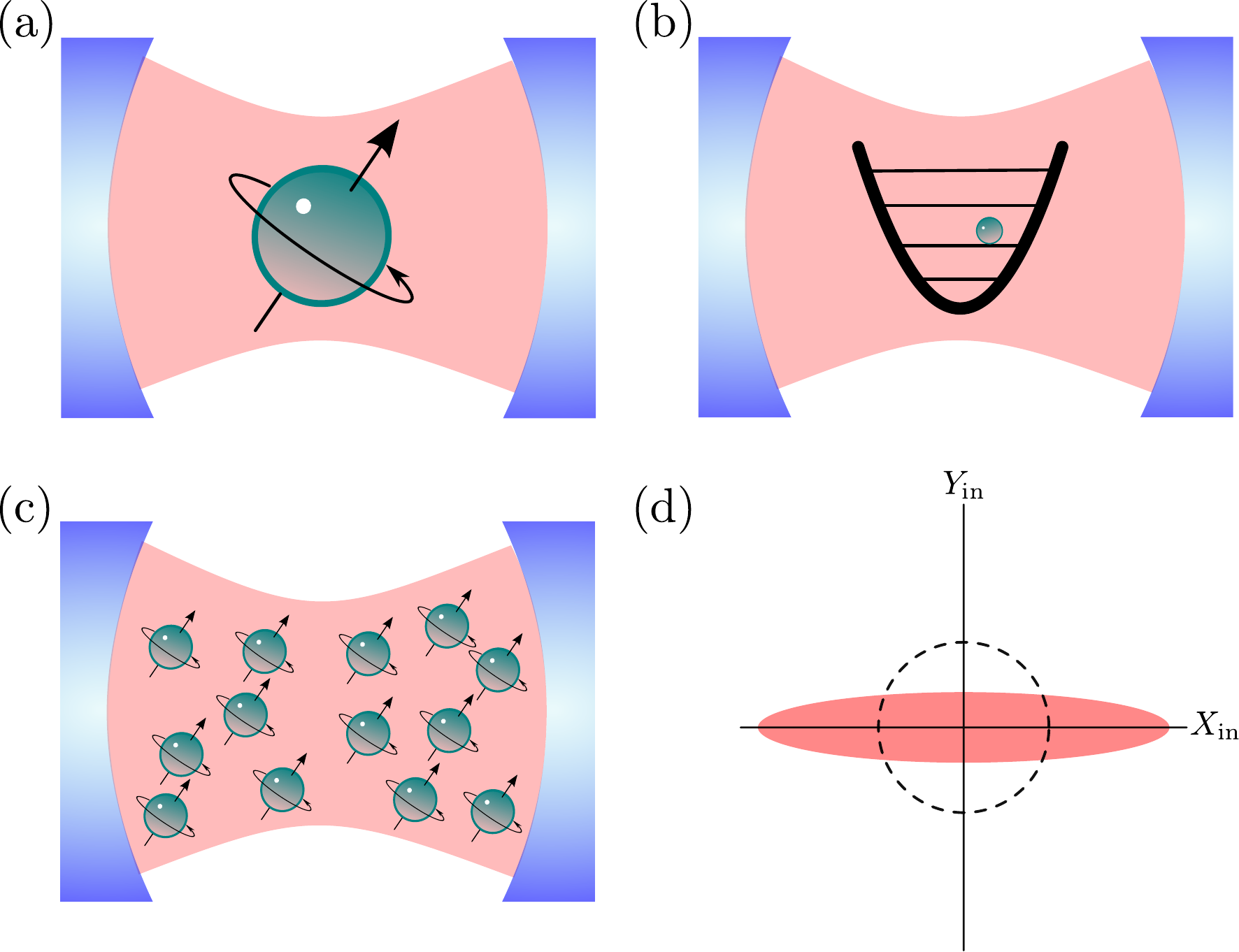}
\caption{A bad cavity is coupled linearly to (a) a single spin-1/2 particle, (b) an oscillator degree of freedom, or (c) a collection of $n$ spin-1/2 particles.
(d) Quadrature representation of the squeezed vacuum input field to the cavity, \eqref{eq:input} with squeezing parameters $N = 1$ and $M=\sqrt{2}$, shown as the red ellipse. The fluctuations in the $\Ynh_{\mathrm{in}}$ quadrature are reduced below those of a vacuum state (dashed circle) at the cost of increased fluctuations in the $\Xnh_{\mathrm{in}}$ quadrature.
}
\label{fig:core}
\end{figure}

In this work we study the dynamics of spin systems due to their coupling to a squeezed bath. As our model system we consider a cavity subject to a squeezed vacuum input and containing either an oscillator or a system of spins to which it is linearly coupled (see \fref{fig:core}). Assuming the bad cavity limit, we eliminate the cavity mode and derive equations of motion for the first and second moments of the oscillator or collective spin components.
Note that in the present study, the systems only couple collectively to the
radiation field and experience no individual damping, which would make the collective system
explore mixed states beyond the large spin Dicke states \cite{PhysRevA.87.062101}. 
We reproduce the result of Gardiner for a single spin, and we show how an oscillator description of the ensemble becomes valid in the limit of many weakly excited spins. Our analysis unveils how the two limiting cases differ by the relative contributions of quantum (vacuum) fluctuation and of radiation reaction to the dynamics.

In Sec.~\ref{sec:model} we give a detailed description of our model and derive Heisenberg equations of motion for the quantum systems of interest. In Sec.~\ref{sec:resDis} we present our main results regarding decay rates and noise fluctuations and we discuss the results.
In Sec.~\ref{sec:conclusion} we conclude and compare the achievements of our formalism to a master equation treatment. We then discuss our results and their applications in a broader context.

\section{Model}\label{sec:model}
We consider a generic model of a quantum system coupled resonantly with strength $g$ to a single cavity mode with field annihilation operator $\ac$.
The cavity serves the purpose of mediating and enhancing the coupling of the system to an input field $\b(t)$ incident on the cavity mirrors \cite{PhysRev.69.37}. We focus on the case where the input is a
broadband squeezed vacuum field defined by the properties
\begin{align}\label{eq:input}
\begin{split}
\braket{\b(t)} &= 0
\\
\braket{\b^\dagger(t)\b(\tp)} &= N\delta(t-\tp)
\\
\braket{\b(t)\b(\tp)} &= M\delta(t-\tp),
\end{split}
\end{align}
where  $|M| \leq \sqrt{N(N+1)}$.
Such a field can be produced by a parametric amplifier driven in degenerate mode  or by coupling of light to non-linear materials \cite{GardinerBook}.
In the following we assume, without loss of generality, the phases to be aligned such that $M$ is real and positive.
As illustrated in \fref{fig:core}(d), the field quadrature
$\X_{\mathrm{in}} = \b+\bd$
has a higher variance, $\braket{\X_{\mathrm{in}}^2}= 2N+2M+1$, than the vacuum state obtained when $N=M=0$, while the variance $\braket{\Y_{\mathrm{in}}^2}= 2N-2M+1$ in the orthogonal quadrature, $\Y_{\mathrm{in}}=i(\bd-\b)$ is reduced. The input field thereby fulfills Heisenberg's uncertainty relation $\braket{\X_{\mathrm{in}}^2} \braket{\Y_{\mathrm{in}}^2}\geq 1$ with equality for all $N$ if $M = \sqrt{N(N+1)}$, which we assume in the following. Squeezed states have been demonstrated with $N\simeq 4.2$ corresponding to variances reduced by $12.7\,\mathrm{dB}$ below the vacuum level \cite{PhysRevLett.104.251102}.

\begin{figure}
\centering
\includegraphics[trim=0 0 0 0,width=0.65\columnwidth]{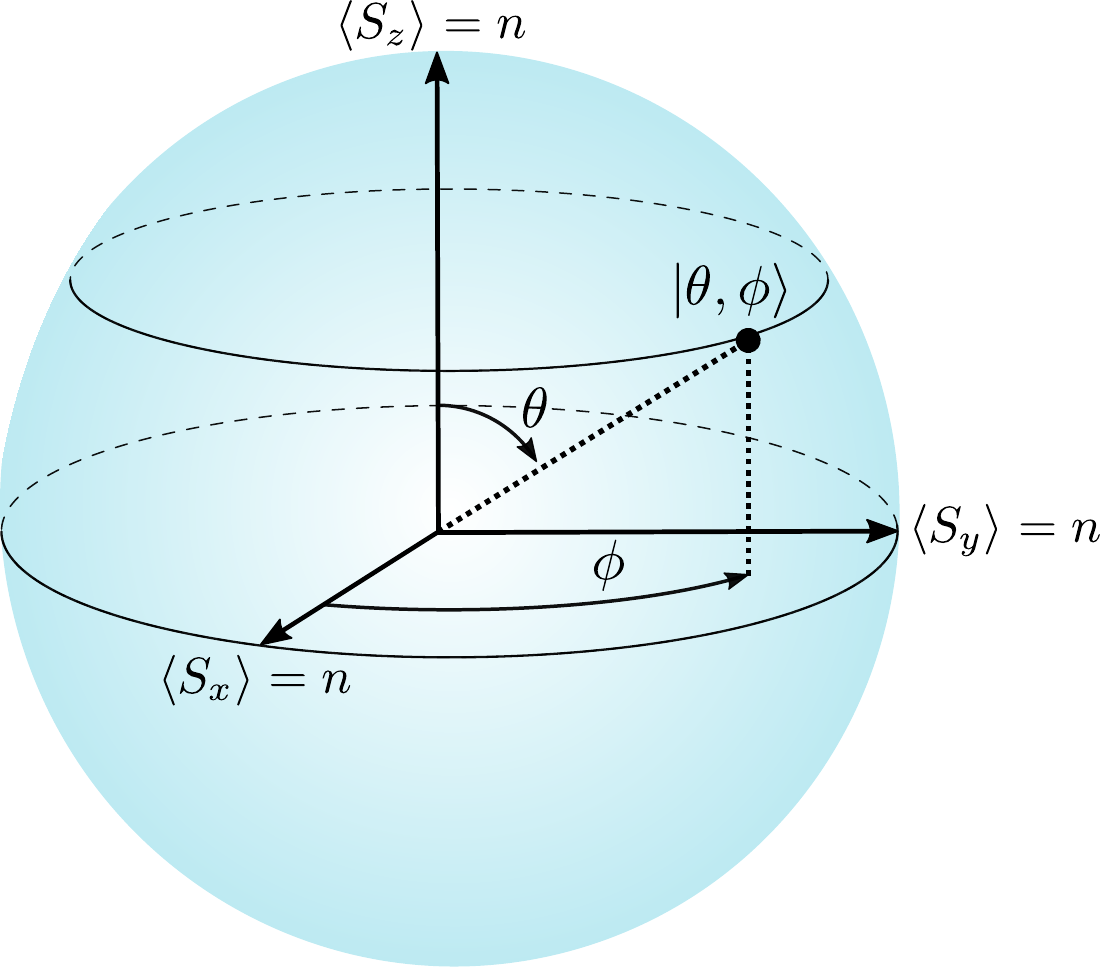}
\caption{Bloch sphere representation of a collective spin with components $(\Sxnh,\Synh,\Sznh)$. The axes go from $-n$ to $n$, the minimum and maximum eigenvalues for an ensemble with $n$ spins.
A spin coherent state $\ket{\theta,\phi}$  is defined in terms of the Bloch angles $\theta$ and $\phi$, [\eqref{eq:spinCoherent}].
}
\label{fig:Bloch}
\end{figure}

\fref{fig:core}(a) shows a single two-level system, represented as a spin-1/2 particle by the Pauli vector of operators $\vec{\sigma} = (\sx,\sy,\sz)$ and the associated lowering and raising operators, $\sm=\sx-i\sy$ and $\sp = \sm^\dagger$, respectively.
\fref{fig:core}(b) shows a harmonic oscillator described by a lowering operator $\a$ with commutator relations $[\a,\ad]=1$ and oscillator quadratures $\X = \a+\ad$ and $\Y=i(\ad-\a)$.
If the single spin is only weakly excited, it behaves as an oscillator since the Bloch vector dynamics described by $(\sx,\sy)$ is similar to the oscillator dynamics described by $(\X,\Y)$ as we heuristically set $\sz=-1$ (remember that $[\sm,\sp] = -\sz$). As quantified in the Holstein-Primakoff approximation introduced below, this similarity is increased as more spins are considered.
To investigate the transition between the single spin and the oscillator results, we hence examine an ensemble of $n$ spins as shown in \fref{fig:core}(c).
The total spin operator is
$\vec{S} = \sum_{i=1}^n\vec{\sigma}^{(i)}$ and the excitation lowering operator is $\Sm = \sum_{i=1}^n\sm^{(i)}$, with $[\Sm,\Sp] = -\Sz$, where $\Sp = \Sm^\dagger$. Notice that, since we omit the factor $1/2$ on the Pauli operators, conventionally used when defining the spin observables, $\vec{S}/2$ ($\vec{S}$) is (is not) an angular momentum.
Symmetric states of the spin ensemble can be represented by an extension of the Bloch sphere picture of a single spin to a collective spin as shown in \fref{fig:Bloch}.

The Holstein-Primakoff transformation maps spin operators to bosonic creation and annihilation operators while conserving the commutation relations \cite{PhysRev.58.1098},
\begin{align}
\begin{split}
\Sz &= 2\ad\a-n,
\\
\Sm &= \sqrt{n-\ad\a}\a,
\\
\Sp &= \ad\sqrt{n-\ad\a}.
\end{split}
\end{align}
For a large ensemble $n\gg 1$ and weak excitation, the second terms in the square roots are negligible such that $\Sm\simeq \sqrt{n}\a$ and $\Sp\simeq \sqrt{n}\ad$. This constitutes the HPA.

\subsection{Heisenberg equations of motion}
The interaction between the system and the cavity field is described by a Tavis-Cummings-type Hamiltonian \cite{PhysRev.170.379} ($\hbar=1$)
\begin{align}\label{eq:H}
\H &= ig\left(\d \acd-\dd \ac\right),
\end{align}
where for a single spin $\d=\sm$, for an ensemble of $n$ spins $\d=\Sm$, and for an oscillator $\d=\a$.
This yields the Heisenberg equation of motion for a general system operator $\q$,
\begin{align}\label{eq:HEoMunSplit1}
\dot{\q} &= g\left([\q,\d] \ac^\dagger - [\q,\dd]\ac\right).
\end{align}
Likewise, if $\kappa$ denotes the output coupling through the cavity mirrors, the quantum Langevin equation for the cavity mode is,
\begin{align}
\dac &= g \d -\frac{\kappa}{2} \ac +\sqrt{\kappa} \b.
\end{align}
In the bad cavity limit ($g\ll\kappa$), we may assume that the cavity relaxes to a steady state faster than the other relevant time scales, i.e., $\dot{\a}_c = 0$, so that
\begin{align}\label{eq:ac}
\ac
&=\frac{2 \b}{\sqrt{\kappa}}+\frac{2g\d}{\kappa}.
\end{align}
This shows how the cavity field is composed of the input field and a field generated by the system inside the cavity.

Although $\a_c$ commutes with any system operator, the individual terms in \eqref{eq:ac} do not. Accordingly, the relative contributions to the evolution of the system operator in \eqref{eq:HEoMunSplit1} depend on the ordering of $\ac^{(\dagger)}$ and the commutators. For a Hermitian operator $\q$, $\dot{\q}$ must also be Hermitian. Dalibard \textit{et. al.} \cite{dalibard1982vacuum} offer the insight that by requiring each part to be separately Hermitian, $\dot{\q}$ can be decomposed unambiguously into two terms with distinct physical meanings. Such splitting has been applied to study the underlying mechanisms in radiative energy corrections and spontaneous emission processes \cite{dalibard1982vacuum,PhysRevA.41.1587}.
Pursuing this line of reasoning, we may apply \eqref{eq:ac} and rewrite \eqref{eq:HEoMunSplit1} as
\begin{align}\label{eq:HEoM}
\dot{\q} &= (\dot{\q})_{\mathrm{ff}}+(\dot{\q})_{\mathrm{sr}},
\end{align}
where
\begin{align}\label{eq:qff}
(\dot{\q})_{\mathrm{ff}} &= \frac{2 g}{\sqrt{\kappa}}\left([\q,\d] \b^\dagger -\b [\q,\dd]\right)
\end{align}
describes interactions with fluctuations in the input field, and
\begin{align}\label{eq:qsr}
(\dot{\q})_{\mathrm{sr}} &= \frac{2g^2}{\kappa}\left([\q,\d]\dd -\d[\q,\dd]\right)
\end{align}
describes electromagnetic self-reaction mediated by the cavity field.


\subsubsection{Oscillator}
We consider first the case where the system is described by an oscillator degree of freedom, $\d = \a$.
Here Eqs.~(\ref{eq:qff}) and (\ref{eq:qsr}) yield for the annihilation operator,
\begin{align}\label{eq:adot}
\begin{split}
(\dot{\a})_{\mathrm{ff}} &= -\frac{2g}{\sqrt{\kappa}}\b,
\\
(\dot{\a})_{\mathrm{sr}} &= -\frac{2g^2}{\kappa}\a.
\end{split}
\end{align}

We thus obtain the equations of motion for the oscillator quadratures
\begin{align}\label{eq:oscEqM}
\begin{split}
\dX &= -\lambda(\b+\bd)-\frac{\gamma_p}{2}\X
\\
\dY &= i\lambda(\b-\bd)-\frac{\gamma_p}{2}\Y,
\end{split}
\end{align}
where we have introduced the Purcell rate $\gamma_p = 4g^2/\kappa$ and an effective coupling constant $\lambda = 2g/\sqrt{\kappa}$.
The field fluctuations, represented by $\b$, drive the fluctuations of the oscillator quadratures while the self-reaction causes their damping.

\subsubsection{Collective spin ensemble}
Consider now the case of an ensemble of $n$ spin-1/2 particles inside the cavity. The ensemble couples linearly to the cavity field via $\d = \Sm$. By Eqs.~(\ref{eq:HEoM})-(\ref{eq:qsr}), the equations of motion for $\Sm$ and $\Sz$ have contributions from both the field fluctuations and self reaction,
\begin{subequations}
\begin{align}
\begin{split}\label{eq:dSa}
(\dSm)_{\mathrm{ff}} &= \frac{2g}{\sqrt{\kappa}}b\Sz,
\\
(\dSz)_{\mathrm{ff}} &= -\frac{4g}{\sqrt{\kappa}}\left(\Sm b^\dagger+b\Sp\right),
\end{split}
\end{align}
\begin{align}
\begin{split}\label{eq:dSb}
(\dSm)_{\mathrm{sr}} &= \frac{2g^2}{\kappa}\Sm\Sz,
\\
(\dSz)_{\mathrm{sr}} &= -\frac{8g^2}{\kappa}\left(\Sm\Sp \right).
\end{split}
\end{align}
\end{subequations}
The operator terms in $(\dSm)_{\mathrm{ff}}$ and $(\dSz)_{\mathrm{ff}}$ give rise to second order processes that contribute  to the same order in $g^2/\kappa$ as the processes in $(\dSm)_{\mathrm{sr}}$ and $(\dSz)_{\mathrm{sr}}$. This follows by a formal integration of the equations over a small time interval $\Delta t\rightarrow dt$,
\begin{widetext}
\begin{align}
\begin{split}
\left[\Sm(t+\Delta t)-\Sm(t)\right]_{\mathrm{ff}} &= \frac{2g}{\sqrt{\kappa}}\int_t^{t+\Delta t}d \tp \, b(\tp)\Sz(\tp),
\\
\left[\Sz(t+\Delta t)-\Sz(t)\right]_{\mathrm{ff}}
&
=\frac{4g}{\sqrt{\kappa}}\int_t^{t+\Delta t}d \tp \,\left\{\Sm(\tp)\b^\dagger(\tp)+\b(\tp)\Sp(\tp)\right\},
\end{split}
\end{align}
and substitution back in Eqs. (\ref{eq:dSa}), keeping only terms to order $g^2/\kappa$,
\begin{align}
\begin{split}
(\dSm)_{\mathrm{ff}} &= \frac{2g}{\sqrt{\kappa}}b\Sz
-\frac{8g^2}{\kappa}\int_t^{t+\Delta t}d \tp \,\left\{b(t)\Sm(\tp)\bd(\tp)+\b(t)\b(\tp)\Sp(\tp)\right\},
\\
(\dSz)_{\mathrm{ff}} &= -\frac{4g}{\sqrt{\kappa}}\left(\Sm b^\dagger+b\Sp\right)
-\frac{8g^2}{\kappa}
\int_t^{t+\Delta t}d \tp \,\left\{\b(\tp)\Sz(\tp)\bd(t)+\b(t)\Sz(\tp)\bd(\tp)\right\}.
\end{split}
\end{align}
Then applying \eqref{eq:input} yields
\begin{align}\label{eq:dSff}
\begin{split}
(\dSm)_{\mathrm{ff}} &= \frac{2g}{\sqrt{\kappa}}b\Sz-\frac{4g^2}{\kappa}\left\{(N+1)\Sm+M\Sp)\right\}
\\
(\dSz)_{\mathrm{ff}} &= -\frac{4g}{\sqrt{\kappa}}\left(\Sm b^\dagger+b\Sp\right)
-\frac{16g^2}{\kappa}(N+1)\Sz,
\end{split}
\end{align}
where we note that contributions by $\delta$ functions evaluated at the lower integral limit are reduced by a factor of 2.

Hence by \eqref{eq:HEoM}, combining Eqs.~(\ref{eq:dSb}) and (\ref{eq:dSff}), we arrive at
\begin{align} \label{eq:eqmotSm}
\dSm &= \frac{\gamma_p}{2}\Sm\Sz+\lambda b\Sz-\gamma_p\left\{(N+1)\Sm+M\Sp)\right\}.
\end{align}
The two transverse spin components are given in terms of $\Sm$ by $\Sx = \Sm+\Sp$ and $\Sy=i(\Sm-\Sp)$. We thus obtain the operator equations of motion for the three spin components,
\begin{align} \label{eq:eqmot}
\begin{split}
\dSx &= \frac{\gamma_p}{2}\left(\Sm\Sz+\Sz\Sp\right)+\lambda\left(\b+\bd\right)\Sz
-
\gamma_p\left\{N+M+1\right\}S_x,
\\
\dSy &= i\frac{\gamma_p}{2}\left(\Sm\Sz-\Sz\Sp\right)+i\lambda\left(\b-\bd\right)\Sz
-
\gamma_p\left\{N-M+1\right\}S_y,
\\
\dSz &= -2\gamma_p\Sm\Sp-2\lambda\left(\Sm b^\dagger+b\Sp\right)
-2\gamma_p(N+1)\Sz.
\end{split}
\end{align}
\end{widetext}
The first term in all three equations is due to the radiation reaction terms and is independent of the properties of the incident noise. The remaining terms are due to the field fluctuations and depend explicitly on the input field properties.

\section{Results}\label{sec:resDis}
In the foregoing section we derived operator equations of motion for the quadratures of an oscillator and the components of a collective spin. Here we will show how these lead to equations of motion for mean values and variances and we will compare and discuss different limits of the theory.

\subsection{Mean values and decay rates}
The mean value equations discussed in this section yield the effective decay rates of the oscillator quadratures and of the spin components.
From Eq.~(\ref{eq:eqmot}) we find, for the three spin components,
\begin{align} \label{eq:MeanEqmot}
\begin{split}
\braket{\dSx} &= \frac{\gamma_p}{2}\braket{\Sm\Sz+\Sz\Sp}
-
\gamma_p\left(N+M+1\right)\braket{S_x}
\\
\braket{\dSy} &= i\frac{\gamma_p}{2}\braket{\Sm\Sz-\Sz\Sp}
-
\gamma_p\left(N-M+1\right)\braket{S_y}
\\
\braket{\dSz} &= -2\gamma_p\braket{\Sm\Sp}-2\gamma_p(N+1)\braket{\Sz}.
\end{split}
\end{align}

\begin{figure*}
\includegraphics[trim=0 0 0 0,width=0.98\textwidth]{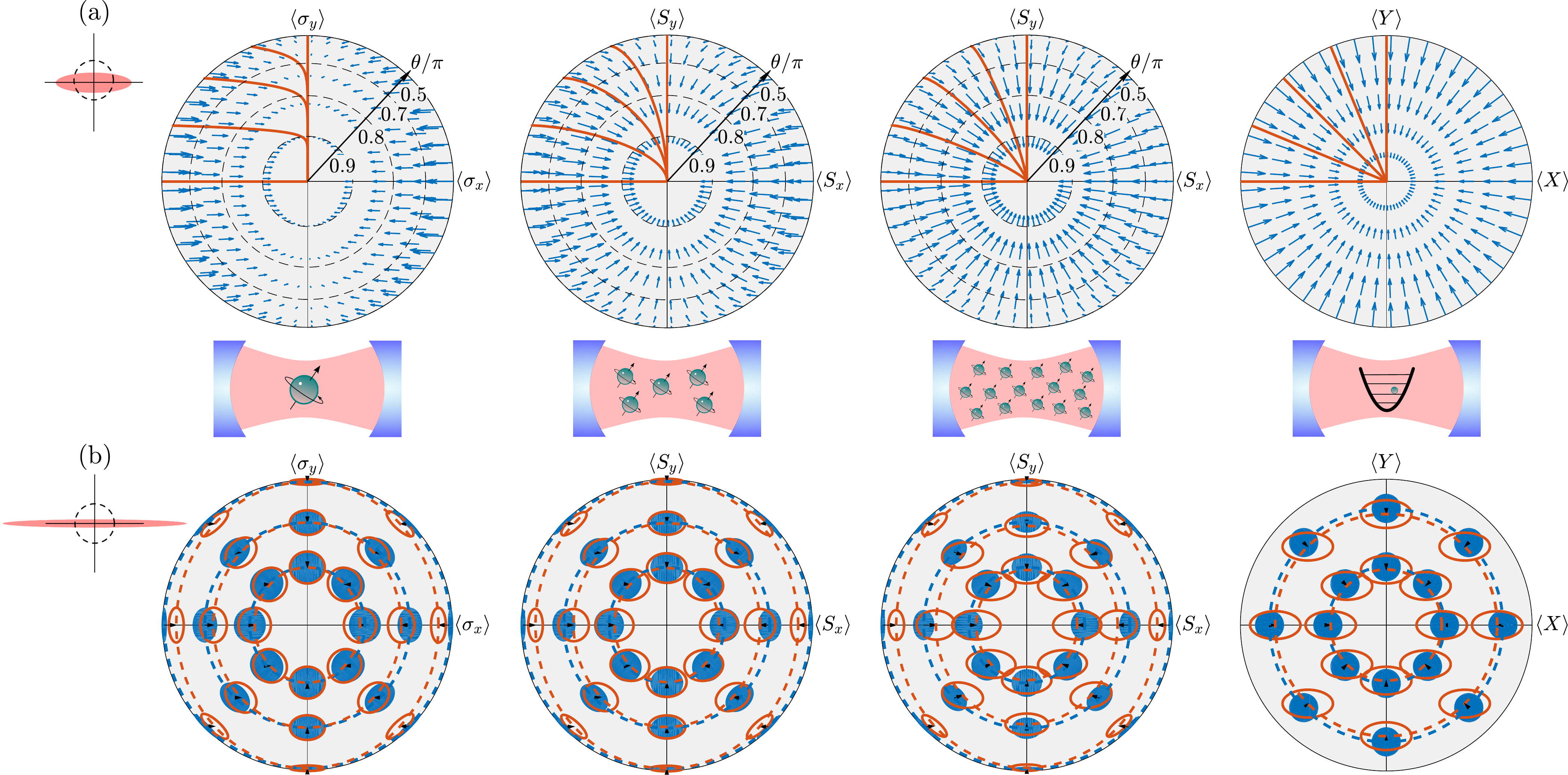}
\caption{\label{fig:combined}
The first three panels in (a) and (b) illustrate the decay of a system with one, five and fifteen spins, respectively, in a squeezed bath with (a) $N=0.5$ and (b) $N=5$.
The plots show the lower half of the Bloch sphere projected on a plane with values of
the polar angle marked by the dashed circles in (a).
(a) Blue arrows indicate the magnitude and direction of the decay of a spin coherent state at each point with the red lines in the second quadrants showing how the collective spin decay progresses from different initial states. The right panel shows a similar plot for the quadrature plane of an oscillator, where the decay is symmetric.
(b) Closed blue circles depict spin coherent states represented by their center coordinates $(\braket{\Sx},\braket{\Sy})$ and error ellipses (see the main text).
States are shown for different $\phi$ and $\theta =0.55,\,0.75,\,0.87\pi$.
For plotting purposes, the error ellipses are scaled by factors $0.12$ for $n=1$ spin, $0.25$ for $n=5$ spins, and $0.4$ for $n = 15$ spins. The mean values and covariance matrix elements are evolved for a short time $\Delta t  = 0.008n\gamma_p^{-1}$ according to Eqs.~(\ref{eq:MeanEqmot}) and(\ref{eq:eqMotVar}). The dashed, blue circle (red ellipse) connects the mean values and the solid blue circles (open red ellipses) show the covariances of the states before (after) the short time evolution.
The right panel shows a similar plot for the oscillator in an initial coherent state evolved for at time $\Delta T = 0.1\gamma_p^{-1}$ by Eqs.~(\ref{eq:meanOsc}) and~(\ref{eq:oscVar}).
}
\label{fig:results}
\end{figure*}

For a single spin, $\sm\sz = \sm$, $\sz\sp=\sp$, and $\sm\sp=(\mathbb{1}-\sz)/2$ and \eqref{eq:MeanEqmot} reduces to the equations for the relaxation of a spin-1/2 particle in a squeezed reservoir derived by  Gardiner \cite{PhysRevLett.56.1917},
\begin{align}\label{eq:meanSingle}
\begin{split}
\braket{\dot{\sigma}_x} &= -\gamma_p\left(N+M+\frac{1}{2}\right)\braket{\sx}
\\
\braket{\dot{\sigma}_y} &= -\gamma_p\left(N-M+\frac{1}{2}\right)\braket{\sx}
\\
\braket{\dot{\sigma}_z} &=
-\gamma_p(2N+1)\braket{\sz}-\gamma_p
.
\end{split}
\end{align}
The relaxation rate of the spin is reduced along the squeezed axis ($y$), while it is increased in the anti-squeezed direction ($x$).

In the case of an oscillator, we have, by Eqs.~(\ref{eq:input}) and~(\ref{eq:oscEqM}),
\begin{align}\label{eq:meanOsc}
\begin{split}
\braket{\dX} &= -\frac{\gamma_p}{2}\braket{\X}
\\
\braket{\dY} &= -\frac{\gamma_p}{2}\braket{\Y}.
\end{split}
\end{align}
The oscillator damping rates are independent of $N$ and $M$ and hence do not depend on the statistics of the input field.

To see how, for a collection of spins where the HPA is valid, the relaxation of the spin degrees of freedom is in agreement with the oscillator damping, we consider the limit of very many spins $n\gg N,M \simeq O(1)$ in \eqref{eq:MeanEqmot}. Close to the south pole of the Bloch sphere, $\Sz\simeq -n$ and we indeed obtain
\begin{align} \label{eq:MeanEqmotLimit}
\begin{split}
\braket{\dSx} &\simeq -\frac{n\gamma_p}{2}\braket{S_x}
\\
\braket{\dSy} &\simeq -\frac{n\gamma_p}{2}\braket{S_y},
\end{split}
\end{align}
equivalent to the mean value equations for oscillator quadratures (\ref{eq:meanOsc}) with $\gamma_p\rightarrow n\gamma_p$. 

To illustrate the transition from the single spin results to the oscillator behavior close to the poles of the Bloch sphere, we consider a spin coherent state. Here all the spins point in the same direction determined by polar angles $(\theta,\phi)$ on the sphere in \fref{fig:Bloch} such that $\braket{\Sx} = n\sin{\theta}\cos \phi$, $\braket{\Sy} = n\sin{\theta}\sin \phi$, and
$\braket{\Sz} = n\cos{\theta}$ with $0\leq\theta\leq\pi$ and $0\leq\phi\leq 2\pi$ \cite{0022-3689-4-3-009},
\begin{align}\label{eq:spinCoherent}
\ket{\theta,\phi} = \bigotimes_{j=1}^{n} \left(\cos \frac{\theta}{2}\ket{\uparrow}_j+\sin\frac{\theta}{2}\e{i\phi}\ket{\downarrow}_j\right).
\end{align}

In such a state, the expectation values in \eqref{eq:MeanEqmot} may readily be calculated, and the effective decay rates of the mean values of the total spin components at each point on the Bloch sphere are determined from our theory.
It yields the evolution
\begin{align}
\begin{split}
\bra{\theta,\phi}\dSx\ket{\theta,\phi} &= -\gamma_x(n,\theta)\bra{\theta,\phi}\Sx\ket{\theta,\phi}
\\
 \bra{\theta,\phi}\dSy\ket{\theta,\phi} &= -\gamma_y(n,\theta)\bra{\theta,\phi}\Sy\ket{\theta,\phi},
\end{split}
\end{align}
where the $n$ and $\theta$-dependent decay rates are
\begin{align}\label{eq:thetaRates}
\begin{split}
 \gamma_x(n,\theta) &= \gamma_p\left\{N+M+\frac{1}{2}-\frac{(n-1)\cos\theta}{2} \right\}
\\
\gamma_y(n,\theta) &= \gamma_p\left\{N-M+\frac{1}{2}-\frac{(n-1)\cos\theta}{2} \right\}.
\end{split}
\end{align}

In the first three panels of \fref{fig:combined}(a), we show the Bloch sphere from below. The dashed circles mark different degrees of excitation, quantified by the polar angle $\theta$.
The arrows indicates the magnitude and direction of the decay of the spin coherent state at each point, and the red lines in the second quadrants guide the eye along the spin decay.
Results are shown for a single spin, five spins and fifteen spins, respectively, and for a moderate degree of squeezing ($N=0.2$).
In the last panel, we show the quadrature plane of an oscillator degree of freedom. Here the arrows and solid lines show the decay of the mean quadratures given by \eqref{eq:meanOsc}.
As predicted, a single spin exhibits a highly asymmetric decay where it quickly relaxes to $\braket{\sy} \simeq 0$, while the $\braket{\sx}$ component decays on a much longer time scale.
The oscillator, on the other hand, shows a completely symmetric decay in the two quadratures.
As the number of spins increases, we see the expected transition between these two extremes, verifying, in this sense, the Holstein-Primakoff approximation.

%
We show in \fref{fig:results2}(a) the relaxation rates of the transverse spin components $\gamma_x(n,\theta)$ and $\gamma_y(n,\theta)$ as functions of $n$ for different values of $\theta$. The results are shown for $N = 0.05$.
As emphasized in \fref{fig:results}, a single spin has highly asymmetric rates but as the number of spins increases, the rates become similar and independent of $N$ and $M$. 
Furthermore, the rates approach those of equivalent oscillator quadratures in the cases where the excitation is very low ($\theta\simeq \pi$).

We can understand the fundamental difference between the single spin and the oscillator cases in terms of contributions to the decay rates from
self-reaction and field fluctuations in \eqref{eq:HEoM}.
For both a weakly excited collective spin and an oscillator, self-reaction processes yield equivalent contributions, $\gamma_p/2$ for the oscillator and $n\gamma_p/2$ for the spins.
Field fluctuations, on the other hand, do not contribute to the oscillator rate, whereas for the spins they cause spontaneous emission that in a squeezed reservoir is adjusted according to the squeezing parameters.
The spin decay by self-reaction is, however, collectively enhanced by the number of spins $n$ such that this contribution dominates the field fluctuations and the Holstein-Primakoff approximation becomes valid when many spins are present.

\subsection{Second moments and Langevin noise}
In the preceding section, we showed that while the statistics of a squeezed reservoir directly influence the decay of a single spin, this effect gradually disappears as more spins are added to the cavity. With many weakly excited spins, the decay resembles that of an oscillator and the mean relaxation is independent of the field statistics.
As shown in \eqref{eq:oscEqM}, however, the input field does enter in the equations of motion for the oscillator quadratures as Langevin noise terms.
Hence, where a squeezed bath influences the decay rates of single spin, it will influence the noise properties of an oscillator degree of freedom.

In this section, we probe these expectations by deriving and investigating equations
of motion for the covariance matrix elements of the transverse spin components and study the limits of a single spin and of an oscillator in the case of low excitation. The non-linearity of a spin system induces non-zero higher moments and we do not expect the covariance matrix to fully characterize the evolution of collective spin states. Nevertheless, we take the (co)variances as indicators for the dynamical evolution of their dominating fluctuations.

The symmetrized covariance between two (noncommuting) observables $A$ and $B$ is defined as
\begin{align}
C_{AB}  &\equiv \frac{1}{2}\braket{\hat{A} \hat{B}+\hat{B}\hat{A}}-\braket{\hat{A}}\braket{\hat{B}}.
\end{align}
If $\hat{A}=\hat{B}$, this is the variance, $V_{A}\equiv C_{AA}$.

The evolution of the (co)variances of the spin components and of the oscillator quadratures can be derived from the equations of motion of the observables in Sec.~(\ref{sec:model}). Since the input field (\ref{eq:input}) is $\delta$ correlated, when taking the time derivative of products of observables, we must employ
Itô's formula \cite{KJstochastic} to obtain
\begin{align}
 \overset{{\boldsymbol{\cdot}}}{\left(\hat{A}\hat{B}\right)}= \dot{\hat{A}}\hat{B} + \hat{A}\dot{\hat{B}}+ \dot{\hat{A}}\dot{\hat{B}}.
\end{align}
With this convention, the evolution of the oscillator quadrature covariance matrix  follows from the equations of motion (\ref{eq:oscEqM}),
\begin{align}\label{eq:oscVar}
\begin{split}
\dot{\text{V}}_{\Xnh} &= -\gamma_p\left\{\text{V}_\Xnh-(2N+2M+1)\right\}
\\
\dot{\text{V}}_{\Ynh} &= -\gamma_p\left\{\text{V}_\Ynh-(2N-2M+1)\right\}
\\
\dot{\text{C}}_{\Xnh\Ynh} &= -\gamma_p\text{C}_{\Xnh\Ynh},
\end{split}
\end{align}
where we applied \eqref{eq:input}.
These expressions clearly show how the oscillator equilibrates with the squeezed input field to a steady state with $V_\Xnh=2N+2M+1$ and $V_\Ynh=2N-2M+1$. Notice that for a coherent state the $+1$ terms in Eqs.~(\ref{eq:oscVar}) ensure that when $N = M = 0$, the variances do not change from $V_{\Xnh}=V_\Ynh = 1$.

We may similarly derive equations of motion for the covariance matrix elements of the $x$ and $y$ components of a collective spin. From \eqref{eq:eqmot} we find after applying commutator relations,
\begin{widetext}
\begin{align}\label{eq:eqMotVar}
\begin{split}
\dot{V}_{\Sxnh} &= -\gamma_p\left\{
(2N+2M+1)\left(V_{\Sxnh}-\braket{\Sz^2}\right)
+\braket{\Sz}
-\C{\Sxnh}{\Sxnh}{\Sznh}
\right\},
\\
\dot{V}_{\Synh} &= -\gamma_p\left\{
(2N-2M+1)\left(V_{\Synh}-\braket{\Sz^2}\right)
+\braket{\Sz}
-\C{\Synh}{\Synh}{\Sznh}
\right\},
\\
\dot{C}_{\Sxnh,\Synh} &= -\gamma_p\left\{(2N+1)C_{\Sxnh,\Synh}-\frac{1}{2}\left(\C{\Sxnh}{\Synh}{\Sznh}+\C{\Synh}{\Sxnh}{\Sznh}\right)\right\}
,
\end{split}
\end{align}
where $\C{A}{B}{C} \equiv \frac{1}{2}\left(\braket{\hat{A}(\hat{B}\hat{C})+(\hat{C}\hat{B})\hat{A}}-\braket{\hat{A}}\braket{\hat{B}\hat{C}+\hat{C}\hat{B}}\right)$.
\end{widetext}
Note that the evolution in general depends on higher moments of the spin observables, signifying how the inherent non-linearity  of spin systems does not preserve Gaussianity.
%

For a single spin, $\hat{\sigma}_i\hat{\sigma}_j=i\epsilon_{ijk}\hat{\sigma}_k$,
where $\epsilon_{ijk}$ is the Levi-Civitá symbol and $i,j,k\in\left\{x,y,z\right\}$, implying that all higher moments are determined by the first moments.
In particular, $V_{\sx}=1-\braket{\sx}^2$ and $V_{\sy}=1-\braket{\sy}^2$, such that, since by \eqref{eq:meanSingle} the transverse spin components relax to zero in the steady state, their variances equilibrate to unity, independent of the squeezed environment. As discussed above, this is opposite to an oscillator degree of freedom.

To investigate how this discrepancy is bridged in the limit of many spins $n\gg N,M$, we consider a spin coherent state for which we note that, e.g., $V_{\Sxnh}$ is proportional to $n$, while $\braket{\Sz^2}$ and $\C{\Sxnh}{\Sxnh}{\Sznh}$ are proportional to $n^2$. Hence, to leading order in $n$ we find
\begin{align}
\begin{split}
\dot{V}_{\Sxnh} &~= \gamma_p\left\{
(2N+2M+1)\braket{\Sz^2}
+\C{\Sxnh}{\Sxnh}{\Sznh}
\right\}
\\
\dot{V}_{\Synh} &~= \gamma_p\left\{
(2N-2M+1)\braket{\Sz^2}
+\C{\Synh}{\Synh}{\Sznh}
\right\}
\\
\dot{C}_{\Sxnh,\Synh} &\underset{n\gg1}{=} \frac{\gamma_p}{2}\left(\C{\Sxnh}{\Synh}{\Sznh}+\C{\Synh}{\Sxnh}{\Sznh}\right)
.
\end{split}
\end{align}
Close to the south pole ($\Sz\simeq -n$), where we expect the HPA to be valid, the equations simplify further,
\begin{align}
\begin{split}
\dot{V}_{\Sxnh} &~= -n\gamma_p\left\{V_{\Sxnh}-
n(2N+2M+1)
\right\},
\\
\dot{V}_{\Synh} &~= -n\gamma_p\left\{V_{\Synh}-
n(2N-2M+1)
\right\},
\\
\dot{C}_{\Sxnh,\Synh} &\underset{\theta\simeq \pi}{=} -n\gamma_p C_{\Sxnh,\Synh}.
\end{split}
\end{align}
I.e., in these limits the noise properties of the collective spin comply with the oscillator results given in \eqref{eq:oscVar}.

The transition from the single spin to an ensemble of many weakly excited spins, is illustrated in \fref{fig:combined}b(). Here we represent spin coherent states (\ref{eq:spinCoherent}) as blue ellipses centered at $(\braket{\Sx},\braket{\Sy})$ with radii along the major and minor axes
given by $(\sqrt{V_{S_a}},\sqrt{V_{S_b}})$, where $V_{S_a}$ and $V_{S_b}$ are the eigenvalues of the covariance matrix.
Upon time evolving the mean values and the (co)variances for a short time $\Delta t\ll \gamma_p^{-1}$ by the expressions given in Eqs.~(\ref{eq:MeanEqmot}) and (\ref{eq:eqMotVar}), the states transform into those shown as open, red ellipses.
The rightmost panel shows a similar plot for an oscillator prepared in a coherent state and evolved under Eqs.~(\ref{eq:meanOsc}) and (\ref{eq:oscVar}).
Dashed lines connecting the $(\braket{\Sx},\braket{\Sy})$ values before and after the time evolution, respectively, show how the states move from a circular to an elliptic configuration, emphasizing the asymmetric rates of the decay.
The uncertainty of a single spin evolves symmetrically in the two transverse spin components while that of an oscillator is squeezed according to the input field. As more spins are added, we see a clear transition between these two limiting cases. The resemblance of the large spin ensemble to the oscillator is more pronounced closer to the center corresponding to low excitation ($\cos\theta\simeq -1$).
\begin{figure}[]
\centering
\includegraphics[trim=0 0 0 0,width=0.95\columnwidth]{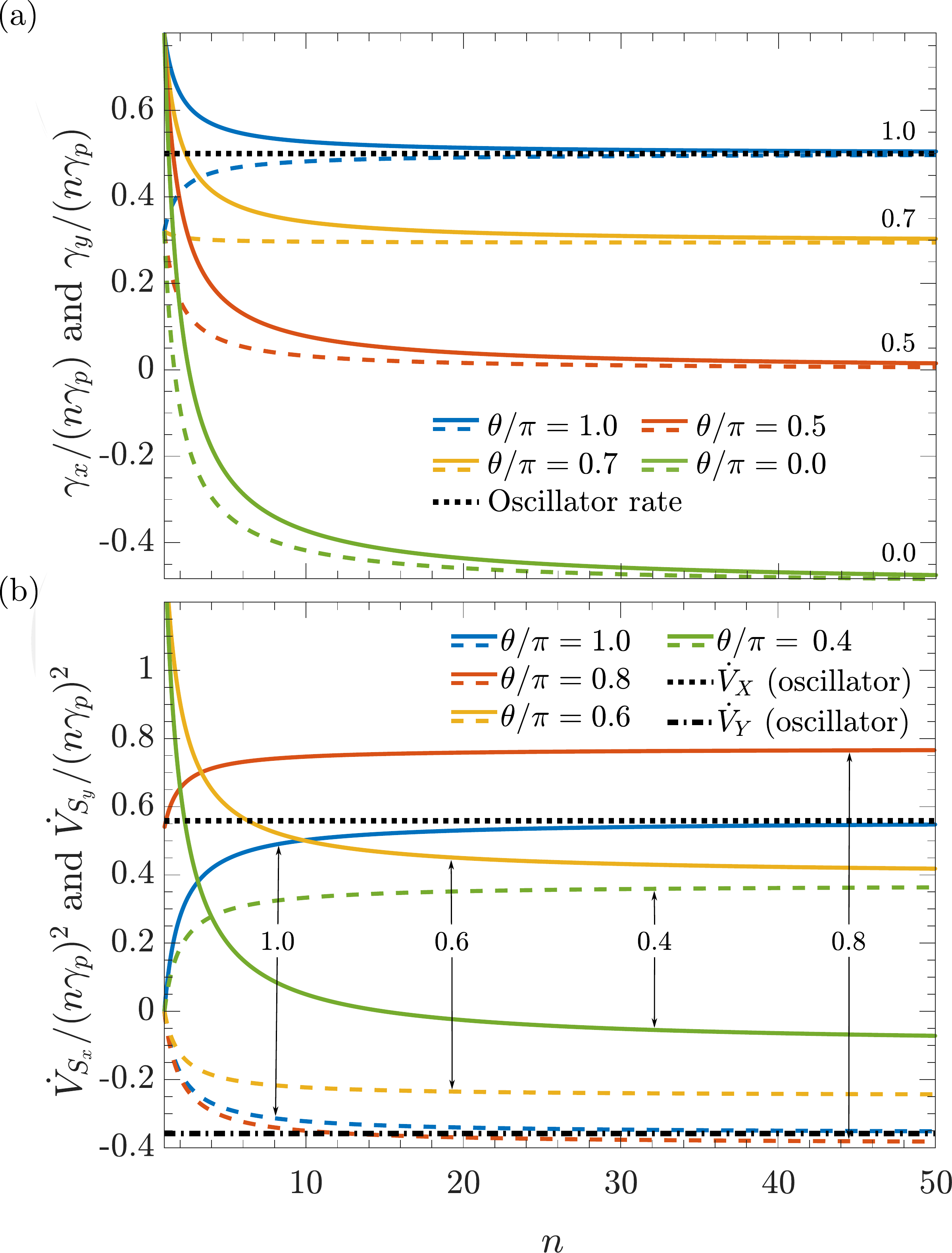}
\caption{\label{fig:4} (a) Decay rates \eqref{eq:thetaRates}
of the $S_x$ (solid lines) and $S_y$ components (dashed lines) of a spin coherent state in a squeezed reservoir with $N=0.05$ shown as a function of the the number of spins $n$ and for different mean excitations quantified by the Bloch angle $\theta$ (values of $\theta/\pi$ are assigned at the right-hand side of the figure). The dotted line shows the symmetric and squeezing independent decay rate $\gamma_p/2$ of either quadrature of an oscillator.
(b) Time derivatives Eqs.~(\ref{eq:eqMotVar}) of variances
of the $S_x$ (solid lines) and $S_y$ (dashed lines) components of a spin coherent state in a squeezed reservoir with $N=0.05$ shown as a function of the the number of spins $n$ and for different $\theta$ with $\phi = 0$ (values of $\theta/\pi$ are assigned by the vertical arrows).
The dotted and dash-dotted lines show the time derivatives of oscillator quadrature variances in similar settings, [Eqs.~(\ref{eq:oscVar})].
}
\label{fig:results2}
\end{figure}

These results are further quantified in \fref{fig:results2}(b), where we compare the time derivatives Eqs.~(\ref{eq:eqMotVar}) of the variances of the two transverse spin components as a function of the number of spins and for different mean excitations $\theta$ to those of oscillator quadratures (\ref{eq:oscVar}).

We note, finally, that our results confirm the findings of Ref.~\cite{Vernac2002} which derives how the squeezing properties of an incoming field can be transferred to a spin ensemble and how this effect increases with the ensemble cooperativity (here equivalent to the number of spins). Reference~\cite{Vernac2002} relies on a linearized model valid only for weak squeezing and considers the steady-state properties of the ensemble, whereas our treatment includes an analysis of the transient behavior and is valid for any amount of squeezing.

\section{Discussion}\label{sec:conclusion}
In this work we have derived equations of motion for the first and second moments of the collective components of a spin ensemble coupled to a broadband squeezed radiation environment. We showed how the effect of the squeezing on the spin dynamics is transferred from the mean values of the components to their fluctuations as more spins are added to the ensemble.
The results thereby link the limiting cases of a single spin, considered by Gardiner \cite{PhysRevLett.56.1917}, and of a large ensemble described well in the Holstein-Primakoff approximation \cite{PhysRev.58.1098}. 

While we formulated our derivations in the Heisenberg picture,
the model described in Sec.~\ref{sec:model} can equivalently be treated in the Schrödinger picture by a master equation of the form \cite{PhysRevA.31.3761},
\begin{align}\label{eq:ME}
\begin{split}
\dot{\rho} &= \frac{\gamma_p}{2}
\Big\{
(N+1)\mathcal{D}[\hat{d}^\dagger,\hat{d}]
+N\mathcal{D}[\hat{d},\hat{d}^\dagger]
-M\mathcal{D}[\hat{d}^\dagger,\hat{d}^\dagger]
\\
&\qquad-M^*\mathcal{D}[\hat{d},\hat{d}]
\Big\}\rho,
\end{split}
\end{align}
where
$\mathcal{D}[\hat{u},\hat{v}]\rho = \hat{v}\rho\hat{u}-\frac{1}{2}\left(\hat{u}\hat{v}\rho+\rho\hat{u}\hat{v}\right)$. In the case of a single spin ($\d=\sm$), the master equation yields ($N,M$)-dependent decay rates of the spin components as given in \eqref{eq:meanSingle}. From the structure of \eqref{eq:ME} it is not apparent that this dependence is lifted once more spins are added.
Explicit calculations, properly taking commutator relations into account, show, however, that the master equation reproduces our results (\ref{eq:MeanEqmot}) for the mean components of a collective spin ($\d=\Sm$), and that in the limit of many weakly excited spins $(\d=\a)$ the squeezing dependent damping terms cancel to realize Eq.~(\ref{eq:meanOsc}).
While the master equation formalism is useful for numerical calculations, it does not offer much insight into the physical mechanisms behind this transition.
Our formalism emphasizes how it occurs as collective radiation reaction processes dominate vacuum fluctuations in the coupling of a large spin system to a squeezed reservoir. Furthermore, the Heisenberg picture treatment allows us to consider second moments for which equations of motion are not easily derived from a master equation.
The results obtained here suggest that the results and interpretations of fluorescence spectra and correlation functions of a spin system in a squeezed bath obtained by a master equation treatment in earlier studies should be revisited \cite{0953-4075-22-1-015,PhysRevA.39.644,CIRAC199026}.

Ensembles of few and many effective spins have been the topic of numerous experimental studies and proposals. To mention a few, their implementations range from storage facilities for microwave excitations \cite{PhysRevLett.105.140503,PhysRevA.85.012333,PhysRevX.4.021049} and optical memories \cite{PhysRevLett.111.020503,PhysRevLett.114.230502,lvovsky2009optical} to quantum information resources \cite{PhysRevLett.83.4204,PhysRevA.95.053819}, gain media in lasers and masers \cite{bohnet2014superradiant,yoshimi2002nuclear}, and to highly sensitive probes \cite{NatureNanotechnology11253257,arXiv:1610.03329,bohnet2014superradiant,Appelaniz2017}. In many of such applications and with the growing ability to engineer squeezing devices in the optical and microwave regimes \cite{andersen201630,PhysRevLett.104.251102,yurke1989squeezing,PhysRevB.89.214517}, the employment of squeezed fields has the potential to enhance readout and performance. For instance, the signal-to-noise ratio in qubit readout can be exponentially increased by utilizing single-mode squeezing \cite{PhysRevLett.115.203601}.
We believe that our simple model and main results, Eqs.~(\ref{eq:MeanEqmot}) and (\ref{eq:eqMotVar}), will be useful in the devising and evaluation of such protocols.
As one example, our results quantify the circumstances under which squeezing of collective spin degrees of freedom \cite{Ma201189} can be achieved by coupling to a squeezed radiation environment, [cf. \eqref{eq:eqMotVar} and \fref{fig:combined}(b)].

There is currently a high interest in superradiance effects and their possible applications \cite{PhysRevLett.95.243602,PhysRevA.81.033847,PhysRevX.6.011025,tieri2017theory}. Superradiance is associated with the enhanced collective radiative rates attained when a significant fraction of the emitters are excited \cite{gross1982superradiance}. This is outside the applicability of the Holstein-Primakoff oscillator approximation but tractable by our analysis.

Besides the insights into the interplay between damping and coherent dynamics in various applications, our study emphasizes a fundamental difference of the role of field fluctuation and radiation reaction between quantum systems with infinite (the oscillator) and low dimensional (single spin-1/2) Hilbert spaces. This difference becomes evident in the interaction with a squeezed environment, and observing the transition between these two extremes as more spins are put into resonance with a cavity may pose an interesting topic for experimental investigation in its own right.


\section{Acknowledgements}
The authors acknowledge financial support from the Villum Foundation.
A. H. K. also acknowledges support from the Danish Ministry of Higher Education and Science.

\bibliography{/home/alexander/Dropbox/PhD/Bibliography/master}{}

\end{document}